\documentclass[twocolumn]{aastex631}
\usepackage{amsmath}
\usepackage{graphicx}
\usepackage{CJK}
\usepackage{booktabs}
\usepackage{hyperref}
\usepackage{xcolor}

\hypersetup{hidelinks,
	colorlinks=true,
	allcolors=blue,
	pdfstartview=Fit,
	breaklinks=true}

\submitjournal{\apj}
\date{\today}

\newcommand\FAST{{\rm FAST}}

\def\src{\rm{PSR J1852$+$0040}}
\def\U4{\rm{4U 0142$+$61}}
\def\SGR{\rm{SGR 0418$+$5729}}
\def\E1{\rm{1E 2259$+$586}}

\begin{document}

\title{Upper limits on the radio pulses from magnetars and a Central Compact Object with FAST}

\correspondingauthor{Ping Zhou}
\email{pingzhou@nju.edu.cn}

\correspondingauthor{Pei Wang}
\email{wangpei@nao.cas.cn}

\author[0000-0001-5653-3787]{Wan-Jin Lu}
\affil{School of Astronomy and Space Science, Nanjing University, Nanjing 210023, China}
\affil{National Astronomical Observatories, Chinese Academy of Sciences, Beijing 100101, China}
\affil{University of Chinese Academy of Sciences, Beijing 100049, China}
\author[0000-0002-5683-822X]{Ping Zhou}
\affil{School of Astronomy and Space Science, Nanjing University, Nanjing 210023, China}
\affil{Key Laboratory of Modern Astronomy and Astrophysics (Nanjing University), Ministry of Education, Nanjing 210023, China}
\author[0000-0002-3386-7159]{Pei Wang}
\affil{National Astronomical Observatories, Chinese Academy of Sciences, Beijing 100101, China}
\author[0000-0001-5684-0103]{Yi-Xuan Shao}
\affil{School of Astronomy and Space Science, Nanjing University, Nanjing 210023, China}
\author[0000-0002-0584-8145]{Xiang-dong Li}
\affil{School of Astronomy and Space Science, Nanjing University, Nanjing 210023, China}
\affil{Key Laboratory of Modern Astronomy and Astrophysics (Nanjing University), Ministry of Education, Nanjing 210023, China}
\author[0000-0002-4708-4219]{Jacco Vink}
\affil{Anton Pannekoek Institute for Astronomy \& GRAPPA, University of Amsterdam, Science Park 904, 1098 XH Amsterdam, The Netherlands}
\author[0000-0003-3010-7661]{Di Li}
\affil{National Astronomical Observatories, Chinese Academy of Sciences, Beijing 100101, China}
\affil{University of Chinese Academy of Sciences, Beijing 100049, China}
\affil{Research Center for Intelligent Computing Platforms, Zhejiang Laboratory, Hangzhou 311100, China}
\author[0000-0002-4753-2798]{Yang Chen}
\affil{School of Astronomy and Space Science, Nanjing University, Nanjing 210023, China}
\affil{Key Laboratory of Modern Astronomy and Astrophysics (Nanjing University), Ministry of Education, Nanjing 210023, China}  

\shorttitle{Radio pulse search of 1 CCO and 3 Magnetars}

\begin{abstract}

Magnetars and central compact objects (CCOs) are subgroups of neutron stars that show a number of properties distinguished from canonical radio pulsars. 
We performed radio observations of three magnetars \SGR, \E1, \U4, and a CCO \src\ with the Five-hundred-meter Aperture Spherical radio Telescope (\FAST) at 1.25 GHz, aiming to search for radio pulsations in their quiescent states. 
During two observation epochs, no radio pulses has been detected towards any target above a significance of signal-to-noise ratio (S/N) = 7 from either the direct folding or blind search. We provided the most stringent upper limit of radio flux ($\lesssim 2$--$4 \ \mu \rm{Jy}$) for the magnetars and the CCO.
For the magnetars with long periods, the real upper limits are likely an order of magnitude larger due to the red noise.
The deep radio observations suggest that these magnetars and the CCO are indeed radio-quiet sources or unfavorably beamed.

\end{abstract}
\keywords{stars: magnetars — stars: neutron — pulsars: individual (\src, \SGR, \U4, \E1)}

\section{Introduction}\label{intro}
Observations of neutron stars (NSs) in the past decades have shown a plethora of classifications, such as traditional rotation-powered pulsars, magnetars, central compact objects (CCOs), and, X-ray dim isolated neutron stars \citep[XDINSs,][]{NSreview18}.

Magnetars are a subgroup of pulsars with ultra-strong magnetic fields (typically  $\gtrsim 10^{14}$ G) and long periods \citep[2--12~s,][]{mereghetti15}.
They are highly variable sources that show high energy bursts, outbursts, giant flares, quasi-periodic oscillations, and sometimes also variable radio emission \citep{review}.  Outbursts of magnetars usually begin with the sudden brightening in the X-ray band and decay slowly in months, which can trigger a diverse emission episode, including periodic radio pulsations.

CCOs are another important subgroup of neutron stars found at the center of the supernova remnants \citep[SNRs,][]{ccoreview}. In contrast to magnetars, spin measurements of three CCOs suggest this subgroup has a weak dipolar magnetic field, which could be explained as anti-magnetars \citep[$10^{10}$--$10^{11}$~G, ][]{antimag08,hg10,gotthelf13},
although there are also alternative scenarios \citep{antimag12}.

So far, magnetars have been primarily discovered from the X-ray bands, while most classical rotation-powered pulsars were identified from the radio pulsations \citep[$\sim 3300$ confirmed\footnote{\href{http://www.atnf.csiro.au/research/pulsar/psrcat}{http://www.atnf.csiro.au/research/pulsar/psrcat}}][]{psrcat}.
Radio emissions have been detected from 6 out of 30 magnetars 
(including 6 candidates) 
following outburst events 
\citep[][]{2006Natur.442..892C, 2007ApJ...666L..93C, 2013Natur.501..391E, 2020ApJ...896L..30E, frb20a, frb20b}{}{}. 
Only the magnetar PSR J1622$-$4950 has been proposed to emit highly variable radio pulsations in the X-ray quiescent phase \citep{PSR4950}.
However, the follow-up X-ray study found that its X-ray flux continued to drop until 2011,  suggesting that the radio activities still occurred during the decay of the X-ray outburst \citep{4950xray12}.

CCOs are also discovered from X-ray observations and, in general, lack of emission in the other wavebands \citep{ccoreview}.
The pulsed emission of magnetars, particularly bright at X-ray and soft $\gamma$-ray band, is in general believed to be powered by the decay of the strong magnetic fields \citep{radtheory1992}, while the X-rays of CCOs are blackbody emission from hot spots on the surface of neutron stars \citep{ccodiscovery03}. 

Whether magnetars and CCOs can generate persisent pulsed radio emissions or single pulses during the quiescence has been an unsolved question.
In the classical paradigm, the beamed radio emission of pulsars is powered by rotational energy losses.
The rotation of the magnetized neutron star induces an enormous electric field, where particles are accelerated to relativistic energies. The subsequent cascade process produces electron-positron pairs (secondary particles) that radiate curvature emission in the radio band \citep{ruderman75}. Old pulsars with low electric potential can not support pair production and thus turn off their radio emission. 
As young pulsars, magnetars and CCOs should also emit radio pulsations if they share the magnetosphere geometry and radiation mechanism with normal pulsars.
However, radio observations in past decades have not  firmly detected any CCO or magnetar during the quiescent state \citep[e.g.,][]{ccolimit1,hg10,ccolimit2,upperlimit}. 
With it comes whether these peculiar NSs are indeed radio-quiet or simply radio-faint. 

Recently, the first Galactic Fast Radio Burst FRB 20200428 was detected from the magnetar SGR~1935+2154 \citep{frb20a,frb20b}, establishing that magnetars are one channel to produce FRBs. 
Radio pulses with a large range of strengths were found from this magnetar, FRBs being just the high fluency outliers \citep{2020ATel14080....1P,kirsten21,zhuww23}.
It is conceivable that other magnetars may also emit FRBs. 
Expanding the single pulse searches in all magnetars and other peculiar pulsars will help in understanding the radio activities and FRB properties of pulsars.
This motivates us to search for single pulses from magnetars and CCO.

We have performed a search of both periodic radio pulsations and single pulses in 3 magnetars, \SGR, \E1, \U4\ and a CCO \src\ in the northern sky using the Five-hundred-meter Aperture Spherical radio Telescope \cite[\FAST\footnote{\href{https://fast.bao.ac.cn}{https://fast.bao.ac.cn}}, see][]{nan11,ieee18}, aiming to find potential faint radio pulses or provide the most stringent constraints on the upper limit if not detected.
We describe the observations of four sources in Section~\ref{sec:Data}. 
The results of the radio pulse searches are introduced in Section~\ref{sec:Result}.
In Section~\ref{sec:Discussion}, we discuss the results briefly and make a conclusion of this work in Section~\ref{sec:Conclusion}.

\section{Observations} \label{sec:Data}

\begin{table}[hb]
\begin{center}
\caption{Observation information of 4 targets}
\resizebox{1.\linewidth}{!}{
\hspace*{-2cm}\begin{tabular}{llccccc}
\hline
\hline
Object & RA & DEC & Date & Epoch &\\
& (h:m:s) & ($^\circ\ ' \ ")$ & (UT) & (MJD) & \\
\hline
\hline
{\it \src} & $18^{h}52^{m}38^{s}.57$ & $+00^\circ 40^{'}19^{"}.8$ & 2019 Jun 21 & $58655.705$ &\\
{\it \SGR} & $04^{h}18^{m}33^{s}.87$ & $+57^\circ 43^{'}23^{"}.9$ & 2019 Jun 21 & $58655.078$ &\\
{\it \E1} & $23^{h}01^{m}08^{s}.30$ & 
$+58^\circ 52^{'}44^{"}.5$ & 2019 Apr 25 & $58598.058$ &  \\
{\it \U4} & $01^{h}46^{m}22^{s}.41$ & 
$+61^\circ 45^{'}03^{"}.2$ & 2019 Apr 25 & $58598.127$ &\\
\hline
\end{tabular}
}
\label{timetable}
\tablecomments{Coordinates and date of FAST observations.}
\end{center}
\end{table}

We conducted \FAST\ observations of \E1, \U4, \src, and \SGR\ on April 25 and June 21, 2019 (obs. ID: 2019-059-P; PI: X.-D.\ Li). 
Integration time for each of these four \FAST\ observations was one hour.
The observation was made in the Pulsar Search Tracking Mode, using the central beam of the 19-beam receiver\citep{jiang20,ieee18} with a frequency coverage of 1.05--1.45 GHz and 3280 channels (122 kHz per channel). 
The data were 8-bit sampled every 49.152 $\mu$s, along with full Stokes polarization parameters recorded.
The noise diode calibration was used at the beginning of the two observations on April 25, 2019. 
No outbursts or other transient signal was reported by high-energy instruments during the observation epoch, which implied that these four sources were at their quiescence. 
The detailed observation information of each source is listed in Table~\ref{timetable}.

\section{Results}\label{sec:Result}

\subsection{Periodicity and single pulse search}\label{sec:search}

We searched for both periodic radio pulsations and single pulses using the software $\tt PRESTO$ \citep{presto}, applying direct folding search, Fourier-Domain blind search with acceleration, and single pulse search \citep{ransom02}, respectively. 
Radio Frequency Interference (RFI) was removed using the command $\tt RFIFIND$ provided by $\tt PRESTO$. 

Before searching, we conducted a de-dispersion of the data by considering a broad range of Dispersion Measures (DM).
The used DM range of \src\ is  0--3000 pc ${\rm cm}^{-3}$, the upper limit of which is several times larger than the value predicted by NE2001 model \citep[][440 pc ${\rm cm}^{-3}$]{CL00}, and three times the value predicted by the Galactic electron-density model YMW16 \citep[][982 pc ${\rm cm}^{-3}$]{YMW16}  at the distance of 7.1 kpc \citep{ccodist,zhou16}.  
The DM values of \SGR, \E1\ and \U4\ predicted by NE2001 are 63.52, 99.71, and 103.83 pc ${\rm cm}^{-3}$ at distances of 2, 3.2 and 3.6 kpc, respectively \citep{SGRdist,1Edist,4Udist}.
We performed the de-dispersion ranging from 0 to 535.95~pc ${\rm cm}^{-3}$ for the three magnetars.
We used a fine DM spacing for the low DM and a slightly coarse step for the high DM (d(DM)=0.05--0.2 pc~${\rm cm}^{-3}$) according to the $\tt DDplan.py$ \citep{presto}, which is designed to balance the de-dispersion accuracy and computing efficiency.

\begin{table*}
\begin{center}
\caption{Spin parameters, DM, and the S/N=7  upper limits on the radio pulses of the four targets}
\resizebox{1\textwidth}{!}{
\hspace*{-1.5cm}\begin{tabular}{l|ccc|ccccccc}
\hline
\hline
Object & $P$  & $\dot{P}$ & DM & $S_{\rm min}$ & $S_{\rm SP,30}$ & $S_{\rm SP,1}$ & $S_{\rm SP,0.1}$ & $S_{\rm min,r}$\\
& (s) & $(10^{-15}  {\rm s~s}^{-1})$ & ($\rm pc\ {\rm cm}^{-3}$) & $(\mu {\rm Jy}$) & ($\rm \ mJy$) & ($\rm \ mJy$) & ($\rm \ mJy$) & ($\rm \ \mu Jy$) &\\
\hline
{\it \src} &  $0.10491264810(3)$ & $0.00868(9)$ &  $440$ &  $2.9$ & 2.5 & 44.5 & 443.8 & 2.9 &\\
{\it \SGR} &  $9.0783869(3)$  & $4(1)$ & $63.52$  & $2.3$ & 2.4 & 13.3 & 45.2 & 32.4\\
{\it \E1} &  $6.9789184643(98)$  & $471.23(3)$ & $99.71$ &  $ 3.8^{+0.8}_{-0.8} $ & $4.0^{+0.8}_{-0.8}$  & $21.6^{+4.4}_{-4.4}$ & $75.8^{+15.7}_{-15.2}$ & $41.8^{+8.8}_{-8.8}$\\
{\it \U4} &  $8.6885092(97)$ & $199(4)$ & $103.83$ & $ 3.5^{+0.7}_{-0.6} $ & $3.7^{+0.7}_{-0.6}$  & $20.2^{+3.7}_{-3.4}$ & $70.8^{+13.0}_{-12.0}$ & $38.4^{+7.6}_{-6.6}$\\
\hline
\end{tabular}
}\label{table:info}

\tablecomments{Period $P$ and period derivative $\dot{P}$ of all the four targets were converted to the observation date using the X-ray spin parameters \citep{hg10,Rea13,dk14}. The DM values of these four sources were estimated using the Galactic electronic-density model NE2001 \citep{CL00}.
$S_{\rm min}$ is the upper limit of periodic pulsations at the central frequency $\nu_{\rm c}=$1.25 GHz.
 $S_{\rm SP}$ is the upper limit of the single pulse, and the subscripts 30, 1, and 0.1 denote a pulse width of 30 ms, 1 ms, and 0.1 ms, respectively.
The 1-$\sigma$ error bars are provided for \E1\ and \U4\ based on the  noise diode calibration.
$S_{\rm min,r}$ is the upper limit of periodic pulses by considering the frequency-dependent noise. \\
}

\end{center}
\end{table*}

\begin{figure}[htbp]
\centerline{
\hfill
\includegraphics[width=0.9\linewidth,angle=270]{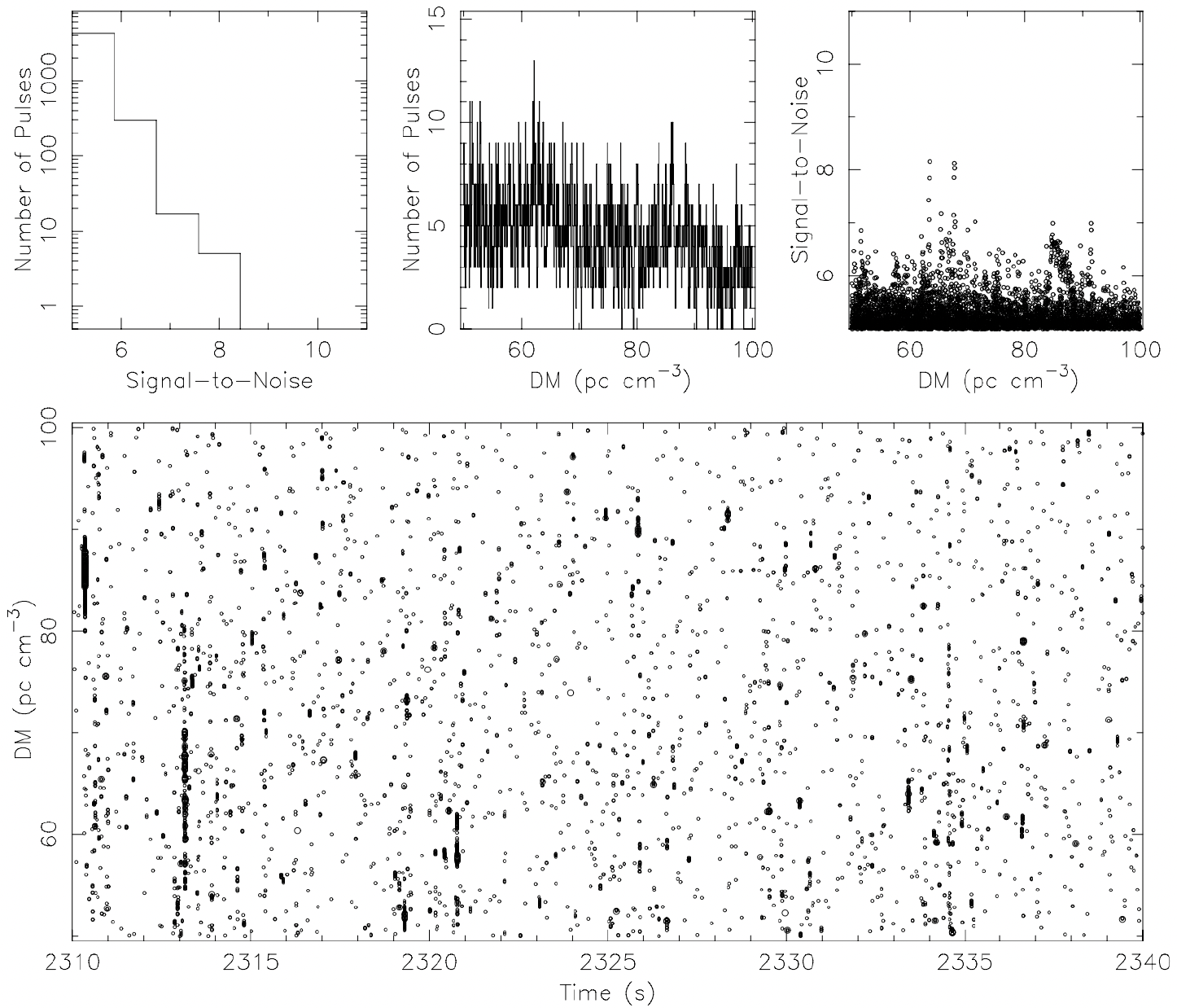}
\hfill
}
\caption{An exemplified image of single pulse search in the DM range 50--100 pc ${\rm cm}^{-3}$, 2310 -- 2340 s of the observation of \SGR. 
The subplot on the top-right could be used to identify the single-pulse candidates. 
The radius of the black dots in the bottom subplot is proportional to its S/N ratio. 
Single pulses will be identified as bunched hollow circles. }
\label{SPplot}
\end{figure}

Since former X-ray observations had precisely measured the spin parameters of the three magnetars and the CCO \citep{dk14,Xraycco,SGRdist,Rea13}, we extrapolated the period of each source (see Table~\ref{table:info}) by calculating the spin period at the current epoch using the X-ray timing results. We then performed a search for periodic pulsations using a derived spin period at our observing epochs for the CCO and three magnetars in the same DM range and with the same steps as mentioned above. 
For this, we simply extrapolated linearly the spin period and used {\tt prepfold} setting the {\tt -nosearch} parameter.
The extrapolated spin periods at the start of our radio observations are 0.10491264810 s, 9.0783869 s, 9.789184643 s and, 8.6885092 s for \src, \SGR, \E1\ and \U4, respectively.
No periodic radio pulsation was found at a S/N = 7 significance. 

Subsequently, we performed a Fourier-Domain acceleration search \citep{ransom02} for latent period signals from four pulsars. We performed the acceleration search in the Fourier domain by adopting $zmax=0,\ 20$ and the threshold of S/N = 7. Firstly, we set $zmax=0$ to search for the periodic signals (including their harmonics) from these four isolated pulsars. 
We then chose  $zmax=20$ because
a simple extrapolation from the old X-ray ephemeris to recent epochs could yield an imprecise spin period, especially for magnetars with large timing noise. 
We folded the candidates of the CCO and three magnetars by using their spin parameters, setting the same threshold S/N = 7. 
We did not search the derivative of spin period $\dot{P}$, since all of the four pulsars are not found in the binary system, and their spin-down level was negligible during the observation. 

Since the number of periodic candidates whose S/N $>$ 7 is acceptable for computing resources, we folded all of them in order to search for any radio pulses that not only from these four sources but also unknown sources that emit radio periodic pulsations.
The folding search was performed using the $\tt prepfold$ command from the software $\tt PRESTO$ \citep{presto}.
No significant periodic radio pulsation was found at a significance of S/N=7 during the observation campaign.
 
We finally searched for single pulses using the de-dispersed time series \citep{cordes03}. 
We used 14 box-car width match filter grids distributed in logarithmic space from 0.1 ms to 30 ms using $\tt single\_ pulse\_ search.py$ in the software $\tt PRESTO$.
For each pulsar, we grouped all the candidates obtained by the search for a time interval of 30~s and DM interval of 50 pc~${\rm cm}^{-3}$ to produce two-dimensional (2D) DM--time plots that we visually inspected to identify significant pulses. 
Significant candidates should be clustered in a high-S/N range ($>7$) near a certain DM value.
As shown in the 2D plot Figure~\ref{SPplot}, several significant single pulse candidates are found.
We then examined the dynamic spectra of all the significant candidates with peak S/N $>7$. 
After cross-checking the 2D plots covering the interested DMs and the dynamic spectra of significant single pulse candidates, we claim that no astrophysical single pulse was found at a threshold of S/N $=7$.

\subsection{Upper limit of radio periodic emission}\label{cal}

We can constrain the upper limits of the averaged flux density of the four targets using two methods 1) theoretical estimation; and 2) noise diode-injected flux calibration.

Theoretically, the minimal detectable radio flux density $S_{\rm min}$ can be estimated by applying the radiometer equation \citep{lk04}:

\begin{equation}\label{eq(A2.6)_fluxden}
    S_{\rm min} = \frac{(S/ N)_{\rm min}\beta T_{\rm  sys}}{G\sqrt{N_{\rm p}t_{\rm obs}(\frac{\Delta f}{\rm MHz})}}\sqrt{\frac{W_{\rm eff}}{P - W_{\rm eff}}} \ \ \ (\rm mJy)
\end{equation}

where $W_{\rm eff} = \sqrt{W^2_{\rm i}+\tau^2_{\rm scatter}+\Delta t^2_{\rm DM}+dt^2}$ represents the effective width of a pulse, $W_{\rm i}$ is the intrinsic pulse width, $\tau_{\rm scatter}$ is the scattering timescale, and $t_{\rm DM}$ is the uncorrected dispersion delay within a frequency channel, introduced by the free electrons along the line-of-sight. $dt$ is the time resolution of the FAST (see in Section~\ref{sec:Data}).
$\beta$ is the digital correction factor, $T_{\rm sys}$ is the system noise temperature, $G$ is the gain depending on telescope performance and the zenith angle, $\Delta f$ is the observation bandwidth in MHz unit, $N_{\rm p}$ is the number of polarisation channels, and $t_{\rm obs}$ is the integration time.
Under the assumption that the performance of FAST and the impact of the atmosphere at our observing frequency stay stable over time \citep{qian20}, we take the gain factor G = 16.7 $\rm K/Jy$, system temperature $T_{\rm sys} = 25.7$ $\rm K$\ \citep{jiang20}, $\beta = 1.1,\ N_{\rm p}=2$, and the integration time is 1 hour.

\begin{figure}[htbp]
\centering

    \begin{minipage}[h]{1\linewidth}
        \includegraphics[width=1\linewidth,angle=0]{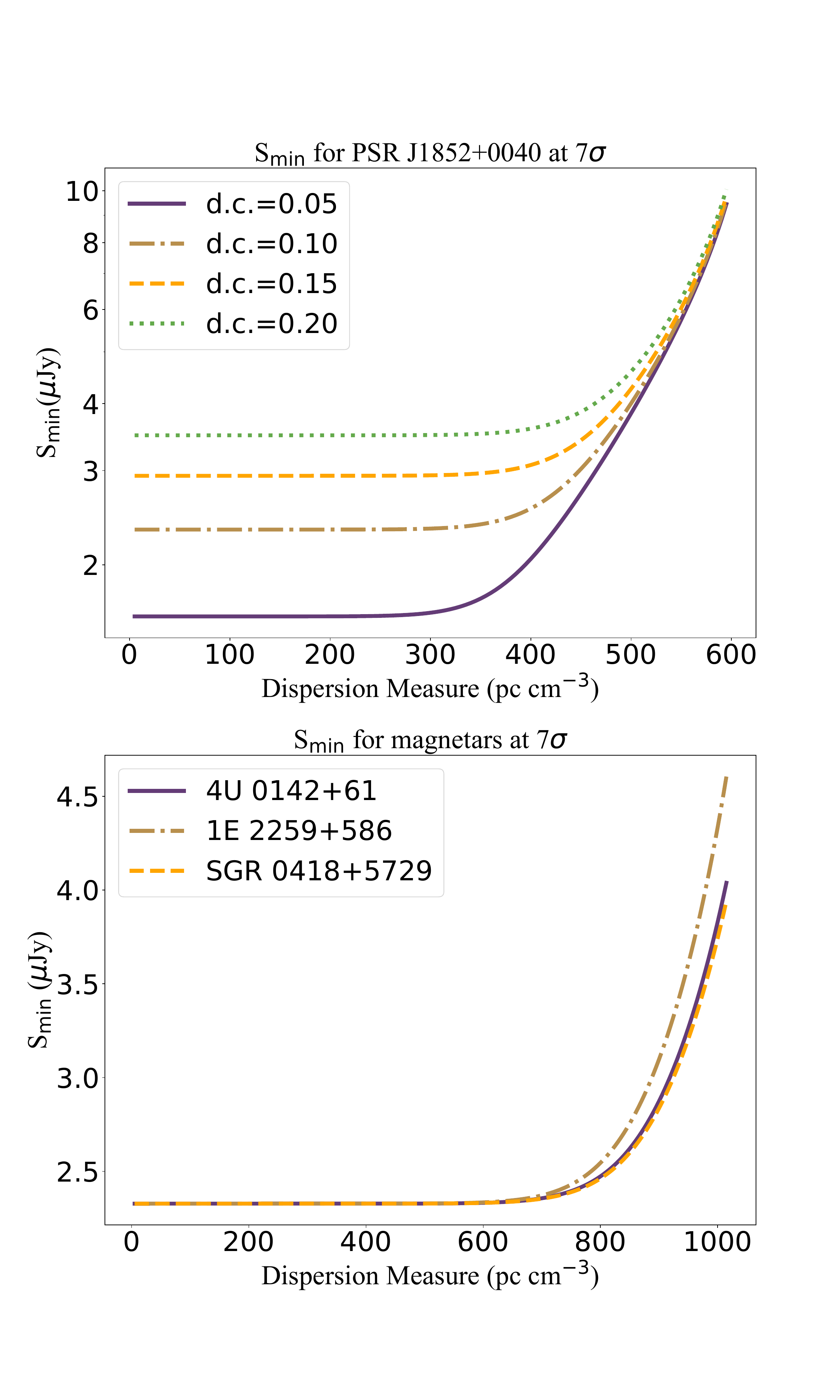}
    \end{minipage}

    \caption{Top: theoretical upper limit of the radio flux density for the CCO \src\ (top panel) varying with the intrinsic duty cycle (d.c.). The flux was cut off at DM  $\sim 640 ~ {\rm pc\ cm}^{-3}$ because of the smearing caused by $\Delta t_{\rm DM}$. We take the DM  $\sim 440 ~{\rm pc\ cm}^{-3}$ predicted by NE2001 model \citep{CL00} to estimate the theoretical upper limit.
    Bottom: theoretical upper limits of radio flux density for magnetars \U4, \E1\ and, \SGR, (assuming an intrinsic duty cycle of 10$\%$).}
\label{Theo:flux}
\end{figure}

The S/N=7 radio flux upper limits for all four sources theoretically estimated by Equation~\ref{eq(A2.6)_fluxden} are 
plotted in Figure~\ref{Theo:flux}.
Neglecting the potential effects of scattering, the figures show that the 
radio flux sensitivity depends mainly on the DM and the intrinsic duty cycle ($\sim W_{\rm i}/P$).
The sensitivity decreases rapidly at high DM values.
This is because the frequency-dependent time delay ($\Delta t_{\rm DM}$) caused by large dispersion would smear the pulse width.
Since this effect shares the inverse square relation with the observing frequency ($\Delta t_{\rm DM} \propto f^{-2}$), observations in higher frequency or with finer frequency resolution can significantly mitigate the pulse smearing and thus, the loss of sensitivity due to dispersion.

    By adopting a S/N=7 significance, the radiometer equation provides the flux upper limits of $2.89 \ \mu \rm{Jy}$ for \src\ and $2.33\ \mu \rm{Jy}$ for three magnetars at the intrinsic duty cycle of 10\%, and assumed DMs as indicated in Table~\ref{table:info}.
These are nearly or more than one order of magnitude lower than the former results observed by GBT \citep{Xraycco,Rea13,upperlimit}.

Besides applying the radiometer equation, a switched calibration noise diode was applied in the first two minutes of observations for \E1\ and \U4. 
This provides a more accurate
system temperature for calculating the upper flux density limits of the observation.
We extracted the calibration data and fitted the baseline and amplitude to calculate the system temperature $T_{\rm sys}$ at the observation date as follows \citep{lk04}

\begin{equation}\label{Tsys}
    \frac{T_{\rm cal}}{T_{\rm sys}} = \frac{\rm OFFCAL-OFF}{\rm OFF}
\end{equation}

where $\rm T_{\rm cal}$ is the equivalent temperature of the noise diode, OFFCAL and OFF represent the counts of system noise plus the noise diode and system noise, respectively.
We considered the uncertainty of DM values predicted by NE2001 \citep{CL00} and the zenith-angle dependence to measure the uncertainty of sensitivity. The zenith-sensitive Gain factor with $1\sigma$ uncertainty was $12.11^{+0.51}_{-0.51}$ K/Jy for \U4\ and $13.60^{+0.14}_{-0.15}$ K/Jy for \E1, respectively. 
We applied these calibrated Gain factors to produce the baseline calibration results for these two sources, which can be found in Table~\ref{table:info} and are summarized in Section~\ref{sec:Conclusion}.

Finally, we noted that the above theoretical calculation holds for ideal white noise conditions. Frequency-dependent noise, such as red noise, can dominate the noise of low-frequency periodic signals  \citep[see][]{rednoise15}. 
To estimate the potential impact of red noise on our sensitivity to periodic pulses from the three magnetars, we calculated the amplitude difference of the noise in the Fourier series between the areas around the spin period of the magnetars and the CCO. We did this by using the RMS power of the neighboring bins of each spin period.

After considering the noise power ratio,
the corrected theoretical upper limit of periodic flux is  $35.4\ \mu \rm Jy$ for \SGR, $27.4^{+5.5}_{-4.4} \  \mu \rm Jy$ for 
\U4, 
and $29.7^{+5.5}_{-5.5}\ \mu \rm Jy$ for \E1.

\subsection{Upper limit of single pulses}\label{sec:sp}
The upper limit of a single pulse can be calculated by \citep{cordes03}\
\begin{equation}\label{fluxSP}
    S_{\rm SP} = \frac{(S/N)_{\rm min}\beta T_{\rm sys}}{{\rm G}W_{\rm i}} \sqrt{\frac{W_{\rm eff}}{N_{\rm p}(\frac{\Delta f}{\rm MHz})}}
\end{equation}
where $W_{\rm eff} $ shares the same definition with Equation~\ref{eq(A2.6)_fluxden} and $W_{\rm i}$ is the intrinsic pulse width.
We also calculated the upper limits of single pulses for all four sources by applying Equation~\ref{fluxSP}.
The instrumental and calibration parameters are the same as used in Section~\ref{cal}.
The results are listed in Table~\ref{table:info}.

\section{Discussion}\label{sec:Discussion}
\begin{figure}[htbp]
\centering
    \begin{minipage}[h]{1\linewidth}
        \includegraphics[width=1.0\linewidth,angle=0]{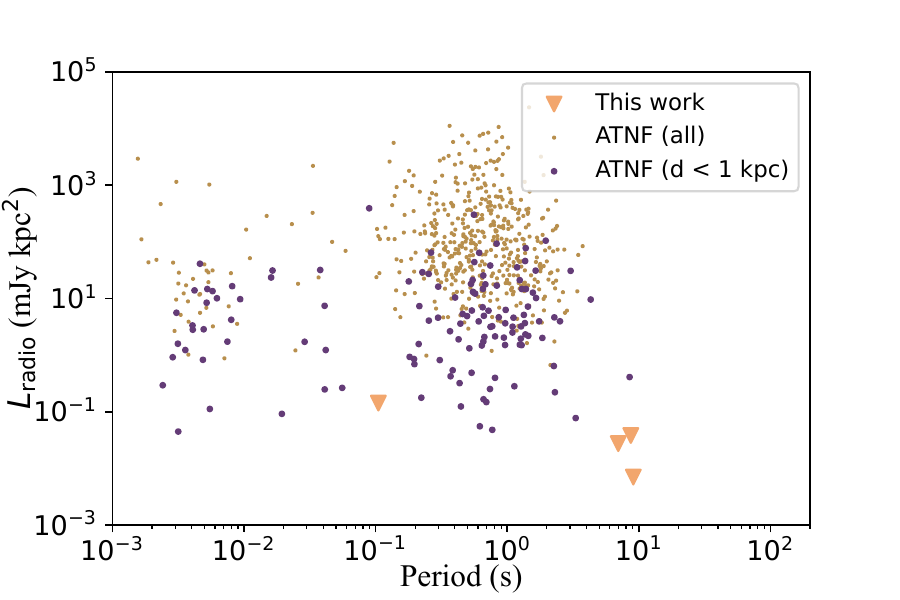}
    \end{minipage}

    \caption{The luminosity of radio pulsars at 1.4 GHz as a function of the spin periods. Purple points represent pulsars with a distance of less than 1 kpc. Orange lower triangles denote the upper limits of periodic pulsation from four pulsars in this work.}
\label{Lum_P}
\end{figure}

Our observations suggest that the CCO and three magnetars are radio-quiet during the observing epochs. 
We compared the upper limits of the pulsed radio luminosity (without the red noise correction for comparison purposes) from the four pulsars with those from known radio pulsars \citep{psrcat}, where we took the luminosity at 1.4~GHz and adopted a typical power-law index $\alpha=-1.6$ for our sources \citep{lori95} to extrapolate from 1.25 GHz to 1.4 GHz. 
As shown in Figure~\ref{Lum_P}, 
the four pulsars in our study are over one order of magnitude dimmer than known radio pulsars with distances larger than 1~kpc.
This reinforces the idea that the three magnetars and the CCO are probably radio-quiet sources.
The upper limits of the four pulsars are comparable or slightly lower than the theoretical estimates made by \cite{wangwy19}, which calculated the coherent radiation from a twisted magnetosphere. 

Here we discuss a few scenarios that have been proposed to explain the non-detection of periodic radio emission in magnetars.

To explain the absence of radio emission in magnetars, \cite{baring98} proposed that the ultra-strong magnetic fields can suppress electron-positron production.
The quantum electrodynamic process, $\gamma \rightarrow \gamma \gamma$ could eliminate photons from pair creation, which is thought to be the power source of radio emission in the polar cap model.

\cite{Rea12} compared the X-ray luminosity and spin-down luminosity of magnetars and high-B pulsars and proposed that the radio emission of magnetars is powered by spin-down energy ($L_{\rm rot}$) that exceeds their quiescent X-ray luminosity $L_{\rm X}$.
{\rm Swift}~J1834.9$-$0846 provided a counter-example to this hypothesis, as its spin-down power is larger than its X-ray luminosity but it is not a radio-loud magnetar \citep{tong13}.
For the radio-loud magnetars mentioned in \cite{Rea12}, the periodic radio emission was discovered during the X-ray/$\gamma$-ray outbursts \citep{review} and faded in several months, which is distinctive from persistent radio pulsations.

\cite{deathline} considered the influence of the Equation of States (EoSs) on a pulsar's death line that distinguishes the radio-loud and radio-quiet regions. For a pulsar with a given inclination angle $\alpha$ between rotation and magnetic axes, the death line is determined by the maximum electric potential $\Phi_{\rm max}\approx \left(\frac{3Ic^3P\dot{P}}{2\pi^2}\right)^{1/2} \ \left(\frac{\Omega^2}{c^2 \sin\alpha }\right)$ \citep{ruderman75,shapiro83}, which depends on the moment of inertia $I$, and thus the equation of state. 
Therefore, the death line in the $P$-$\dot{P}$ diagram can vary with the equation of state.
This might explain the non-detection of radio emission from some magnetars and CCOs, but not all of them.

Besides the aforementioned scenarios, we also considered the beaming effect as a possible reason for the non-detection of the magnetars' radio pulsations. 18 of 24 confirmed magnetars were not reported to emit radio pulsations.
It has been proposed that the radio beaming fraction of the pulsar $f$ is a function of its period \citep{beam}:

\begin{equation}\label{beaming}
    f = 9\left[\rm log\left(\frac{P}{\rm s}\right) - 1\right]^2 + 3\ \ (\%)
\end{equation}

According to the relationship, we obtained a probability of $$\prod_{i=1}^{18} (1-f_{\rm i})=51.9\%$$ when all the 18 confirmed magnetars \citep{magnetarcat} have their radio emission beamed away from the Earth. 
This result implies that there is a probability that these magnetars emit radio pulses but they are unfavorably beamed. 

For CCOs with low dipole magnetic fields (B $\leq 10^{12}\ \rm G$), the lack of radio emission can not be explained with the high magnetic field scenario by \cite{baring98}.
We estimated a non-negligible probability of 31\% when the three CCOs (with spin periods measured from the X-rays) are beamed away in the radio band.
Assuming that all the 10 known CCOs have a period in the range of 0.1--0.4~s \citep{ccoreview}, the missing the emission due to the beaming effect should occur at a small
probability of $ 0.7\% \leq  p \leq 10.0\% $.

The uncertainty of the beaming effect is large, since the spin periods of most CCOs are unknown and thus influence the values of the estimated beaming fraction. 

Therefore, the beaming effect cannot be excluded to explain the non-detection of radio pulses in some magnetars and CCOs. It is also possible that other physical mechanisms, such as the suppression of electron-positron pair production in strong magnetic fields, the energy budget of spin-down luminosity, and the influence of the EoSs, work together with the beaming effect to prevent the detection of persistent radio pulsations from these two subgroups of pulsars.

We have not detected single pulses from the magnetars and the CCO as elaborated in Section~\ref{sec:sp}.
Since the occurrence rate of the radio pulses from these sources is unclear, missing single pulses in our observations does not mean that these sources cannot emit FRBs or other pulses beyond our observation epochs.

\section{Conclusion}\label{sec:Conclusion}

We have performed deep radio searches for pulsations in three magnetars and a CCO and have not found any radio periodic emission or single pulses during the observation epochs. 
The main results are summarized as follows:

\begin{enumerate}

\item The S/N = 7 upper limits of the periodic radio pulsation flux densities have been constrained to be $2.9 \ \mu \rm Jy$ for \src, 
$2.3 \ \mu \rm Jy$ for \SGR, $3.8^{+0.8}_{-0.8} \ \mu \rm Jy$ for \E1\ and $3.5^{+0.7}_{-0.6} \ \mu \rm Jy$ for \U4, by using the radiometer sensitivity equation. These upper limits are almost or more than one order of magnitude lower than earlier results \citep{Rea13,upperlimit,hg10,gaensler01,burgay06} and  comparable with the faintest radio pulsars discovered in the FAST commissioning phase \citep{wang2021SCPMA}.
Considering the frequency-dependent redded noise, the upper limits of the three magnetars were
$32.4 \ \mu \rm Jy$, $41.8^{+8.8}_{-8.8} \ \mu \rm Jy$ and $38.4^{+7.6}_{-6.6} \ \mu \rm Jy$, respectively. 

\item The upper limits of single pulses with a pulse width of 30 ms are $3.7^{+0.7}_{-0.6}\ \rm mJy$ for \U4, $4.0^{+0.8}_{-0.8}\ \rm mJy$ for \E1. 
The theoretical results of \U4\ and \E1\ are nearly one order of magnitude lower than the former study \citep{upperlimit}.
We also provided the very first upper limits of single pulses, $2.4\ \rm mJy$ for \SGR\ and $2.5\ \rm mJy$ for the \src, respectively. 

\item The beaming effect cannot been excluded as a potential reason for the non-detection of pulsations from these magnetars and CCOs. 
Extending the number of deep radio observations on magnetars and CCOs will help to improve our physical interpretation of the missing radio emission in these two subgroups of neutron stars.

\end{enumerate}

\section{Acknowledgments}

We thank the anonymous referee for constructive comments that were helpful in improving the manuscript.
W.-J.L. and P.Z. acknowledge the support from National Science Foundation of China grant No.\ 12273010 and Nederlandse
Organisatie voor Wetenschappelijk Onderzoek (NWO) Veni Fellowship, grant No. 639.041.647. 
P.W. acknowledges support by the NSFC No.\ U2031117 and the Youth Innovation Promotion Association CAS (Grant No. 2021055).
D.Li acknowledges the support by the NSFC No.\ 11988101. 
Y.C. acknowledges the support from NSFC grants under No.\ 12173018 and 12121003.
The computation was made using the facilities at the High-Performance Computing Center of Collaborative Innovation Center of Advanced Microstructures (Nanjing University).


\end{document}